\documentclass[a4paper,twoside,twocolumn]{article}
\pdfoutput=1

\usepackage{
url       %
, fancyhdr % Fancy Headers
, lastpage % Creates marker with key LastPage, for pxxx of yyy to refer to.
, amsmath  % AMS maths
, balance % Balance columns at end of any page with \balance in the left column
}
\usepackage[noindent, arraystretch, fullpage]{setlengths} % [noindent, arraystretch, {fullpage|tightpage|inchrndtext}]{setlengths}
\usepackage{
own         % Defines \newboolean{twocol}
}

\usepackage{ifpdf}  % Avoid \newif\ifpdf which clashes with same command
\ifpdf
    \usepackage[pdftex]{graphicx}
    \usepackage[colorlinks,linkcolor=blue,citecolor=blue,urlcolor=blue]{hyperref}
    \DeclareGraphicsExtensions{.pdf}
\else
    \usepackage{graphicx}
    \usepackage[hypertex]{hyperref}  % supports hypertext in PDF but no Acrobat features e.g. bookmarks
\fi
\usepackage{todonotes} % Allows insertion of \todo{} notes - uses graphicx package and clashes with pdftex option below
\graphicspath{{images/}}

\newcommand*{\metaauthori}{Bob Briscoe}

\newcommand*{\metashorttitle}{Insights from Curvy RED}
\newcommand*{\metatitle}{Insights from Curvy RED (Random Early Detection)}
\newcommand*{\metano}{BT TR-TUB8-2015-003}
\newcommand*{\metakeywords}{Data Communications, Networks, Internet,
Control, Performance, Latency, Responsiveness, Dynamics, Algorithm, Standards}

\newcommand*{\metamaild}{@bobbriscoe.net}
\newcommand*{\metamaili}{\href{mailto:ietf\metamaild}{ietf}}

\newcommand*{\metaaddress}{Formerly of BT Research \& Technology, UK, during the drafting of this report. Now independent.}

\newcommand*{\metaversion}{Issue 01C}
\newcommand*{\metadate}{14 Aug 2015}

\hypersetup{                       % Set PDF document attributes
     pdfauthor = {\metaauthori},
     pdftitle = {\metashorttitle},
     pdfsubject = {},
     pdfkeywords = {\metakeywords}
}%

\title{\metatitle}%
\author{\metaauthori%
\thanks{\metamaili\metamaild, %
\metaaddress}%
}
\date{\metadate}%

\fancypagestyle{first}{%
\fancyhead[LO,RE]{}%
\fancyhead[LE,RO]{}%
\fancyhead[C]{}%
}%

\begin{document}
\bibliographystyle{alpha}%

% ----------------------------------------------------------------

\maketitle%
\thispagestyle{first}

% ----------------------------------------------------------------
\begin{abstract}
{\small\noindent%
% !TeX root = credi_tr.tex
Active queue management (AQM) drops packets early in the growth of a
queue, to prevent a capacity-seeking sender (e.g.\ TCP) from keeping
the buffer full. An AQM can mark instead of dropping packets if they
indicate support for explicit congestion notification
(ECN~\cite{IETF_RFC3168:ECN_IP_TCP}). Two modern AQMs
(PIE~\cite{Pan15:PIE_I-D} and CoDel~\cite{Nichols12:CoDel}) are
designed to keep queuing delay to a target by dropping packets as load
varies.

This memo uses Curvy RED and an idealised but sufficient model of TCP
traffic to explain why attempting to keep delay constant is a bad idea,
because it requires excessively high drop at high loads. This high drop
itself takes over from queuing delay as the dominant cause of delay,
particularly for short flows. At high load, a link is better able to preserve
reasonable performance if the delay target is softened into
a curve rather than a hard cap.

The analysis proves that the same AQM can be deployed in different 
parts of a network whatever the capacity with the same optimal 
configuration.

A surprising corollary of this analysis concerns cases with a highly 
aggregated number of flows through a bottleneck. Although aggregation reduces queue
variation, if the target queuing delay of the AQM at that bottleneck is 
reduced to take advantage of this aggregation, TCP will still increase 
the loss level because of the reduction in round trip time. The way 
to resolve this dilemma is to overprovision (a formula is provided).

Nonetheless, for traffic with ECN enabled, there is no harm in an AQM 
holding queuing delay constant or configuring an AQM to take advantage 
of any reduced delay due to aggregation without over-provisioning. 
Recently, the requirement of the ECN
standard~\cite{IETF_RFC3168:ECN_IP_TCP} that ECN must be treated the
same as drop has been questioned. The insight that the goals of an AQM
for drop and for ECN should be different proves that this doubt is
justified.
}      % Abstract
\end{abstract}
%\input{credi_tr_intro-data}       % Intro
% !TeX root = credi_tr.tex
% ================================================================
\section{Curvy RED}\label{sec:credi_Curvy_RED}

\begin{figure}
  \centering
  \includegraphics[width=\linewidth]{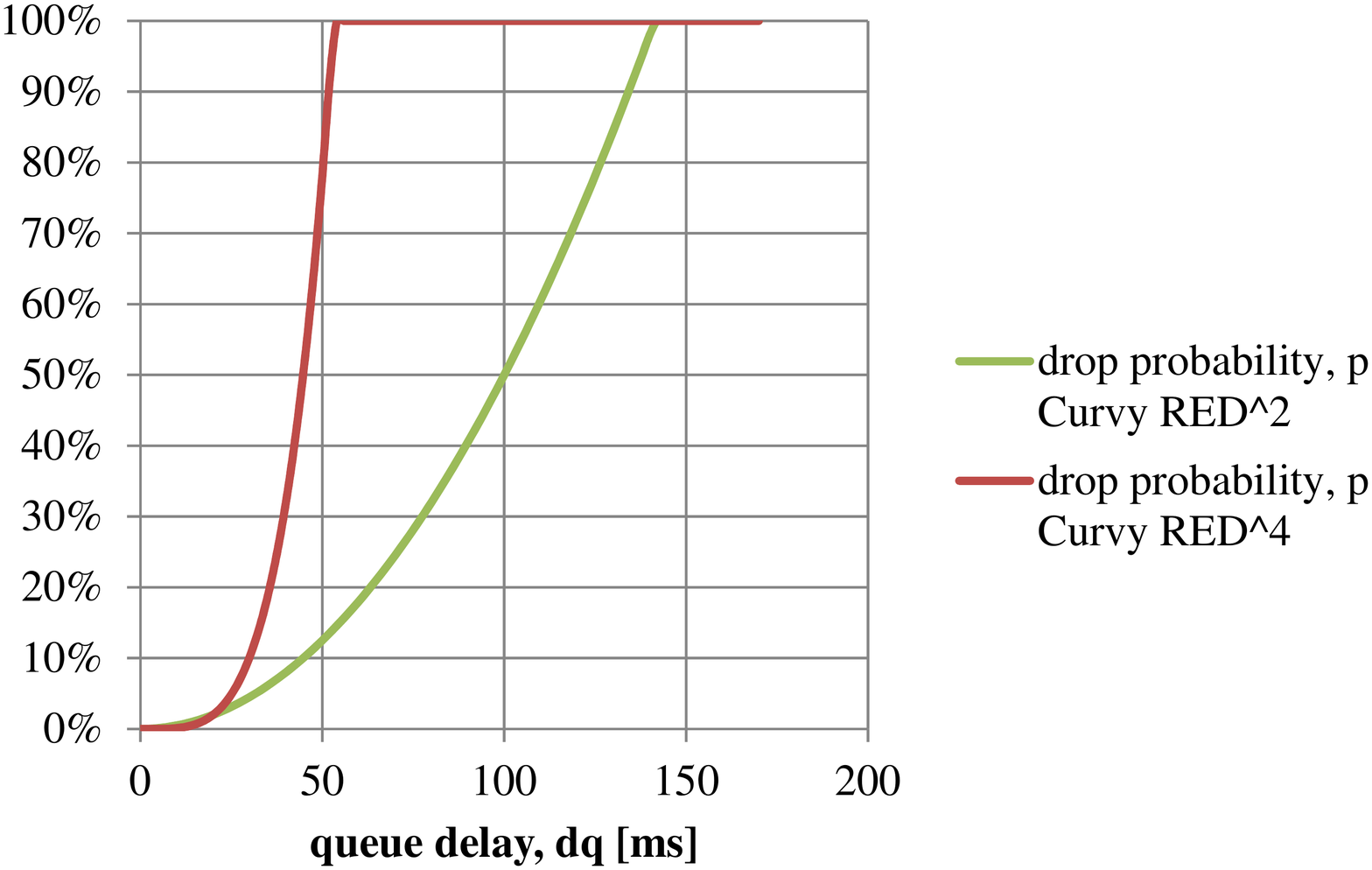}\\
  \caption{Two Example Curvy RED algorithms}\label{fig:credi_curvy-red-100}
\end{figure}

\begin{figure}
  \centering
  \includegraphics[width=\linewidth]{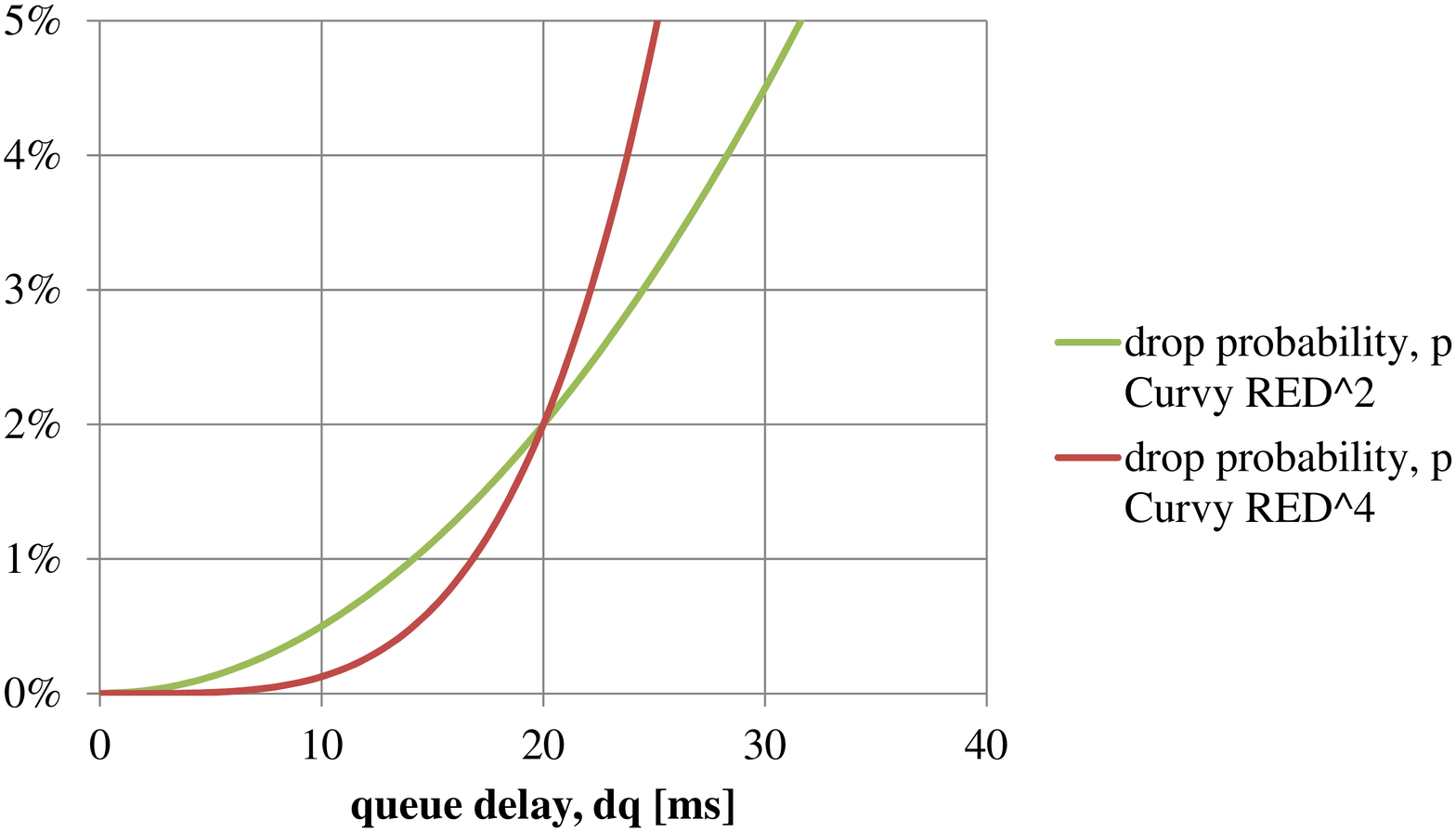}\\
  \caption{Usual Operating Region of The Same Two Example Curvy RED AQM algorithms}\label{fig:credi_curvy-red-5}
\end{figure}

Curvy RED~\cite{DeSchepper15b:DCttH_TR} is an active queue management (AQM) algorithm that is both a
simplification and a a generalisation of Random Early Detection
(RED~\cite{Floyd93:RED}). Two examples are shown in
\autoref{fig:credi_curvy-red-100} and \autoref{fig:credi_curvy-red-5}
shows a close-up of their normal operating regions.

The drop probability, \(p\), of a Curvy RED AQM is:
\begin{equation}
    p = \left(\frac{d_q}{D_q}\right)^{u},
    \label{eqn:credi_cred}
\end{equation}
where \(d_q\) is averaged queue delay. The two parameters are:
\begin{description}
  \item[\(u\):] the exponent (cUrviness) of the AQM;
  \item[\(D_q\):] the scaling parameter, i.e.\ the value of \(d_q\) when \(p\) hits
      100\%. \(1/D_q\) is the slope of the curve.
\end{description}

An implementation would not be expected to vary cUrviness, \(u\). Rather
an optimal value of \(u\) would be determined in advance.

The queuing delay used by Curvy RED is averaged, but averaging is outside 
the scope of his memo, which is only concerned with insights from
steady-state conditions.

\begin{figure}
  a) \includegraphics[width=\linewidth]{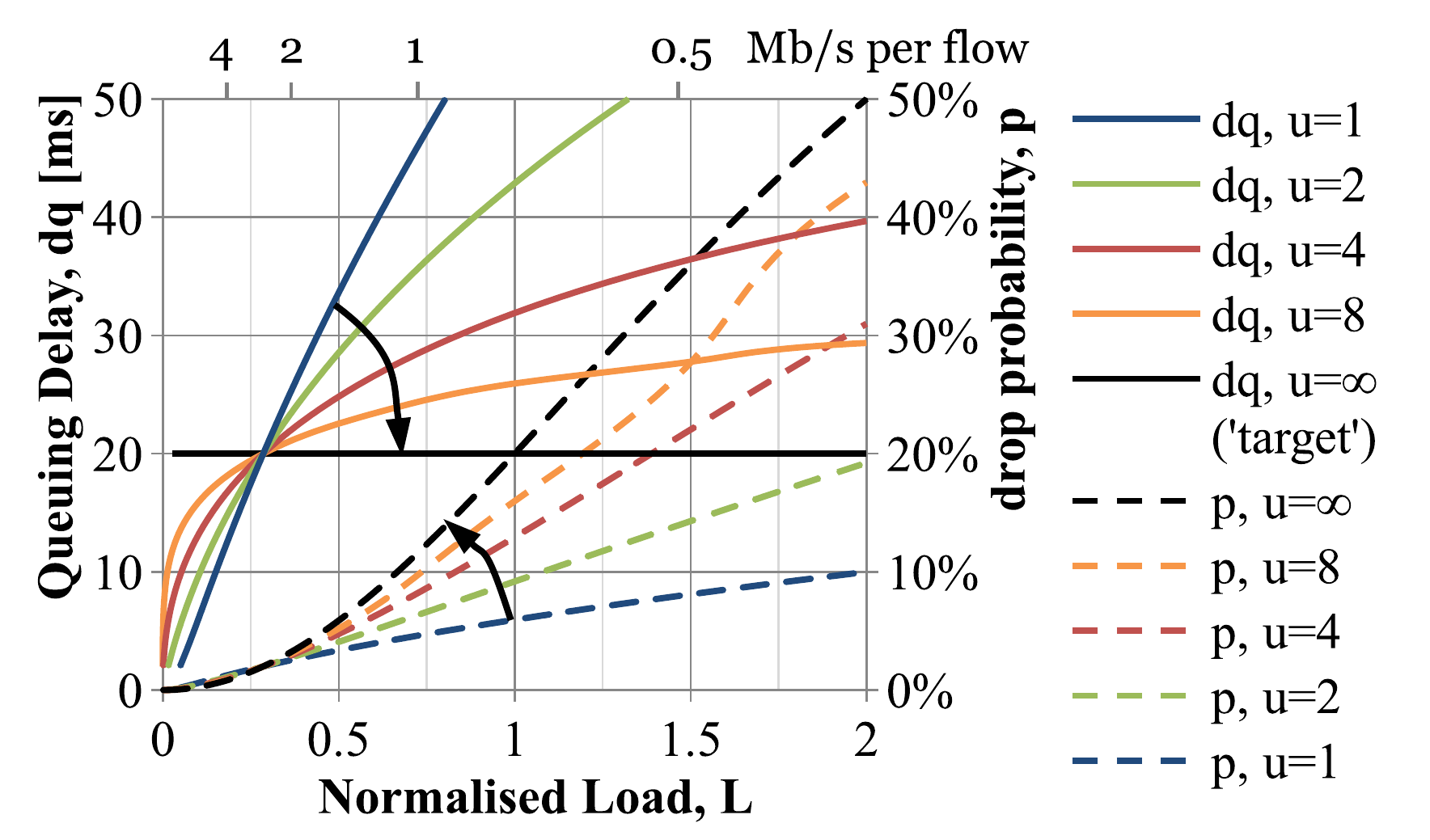}\\
  b) \includegraphics[width=\linewidth]{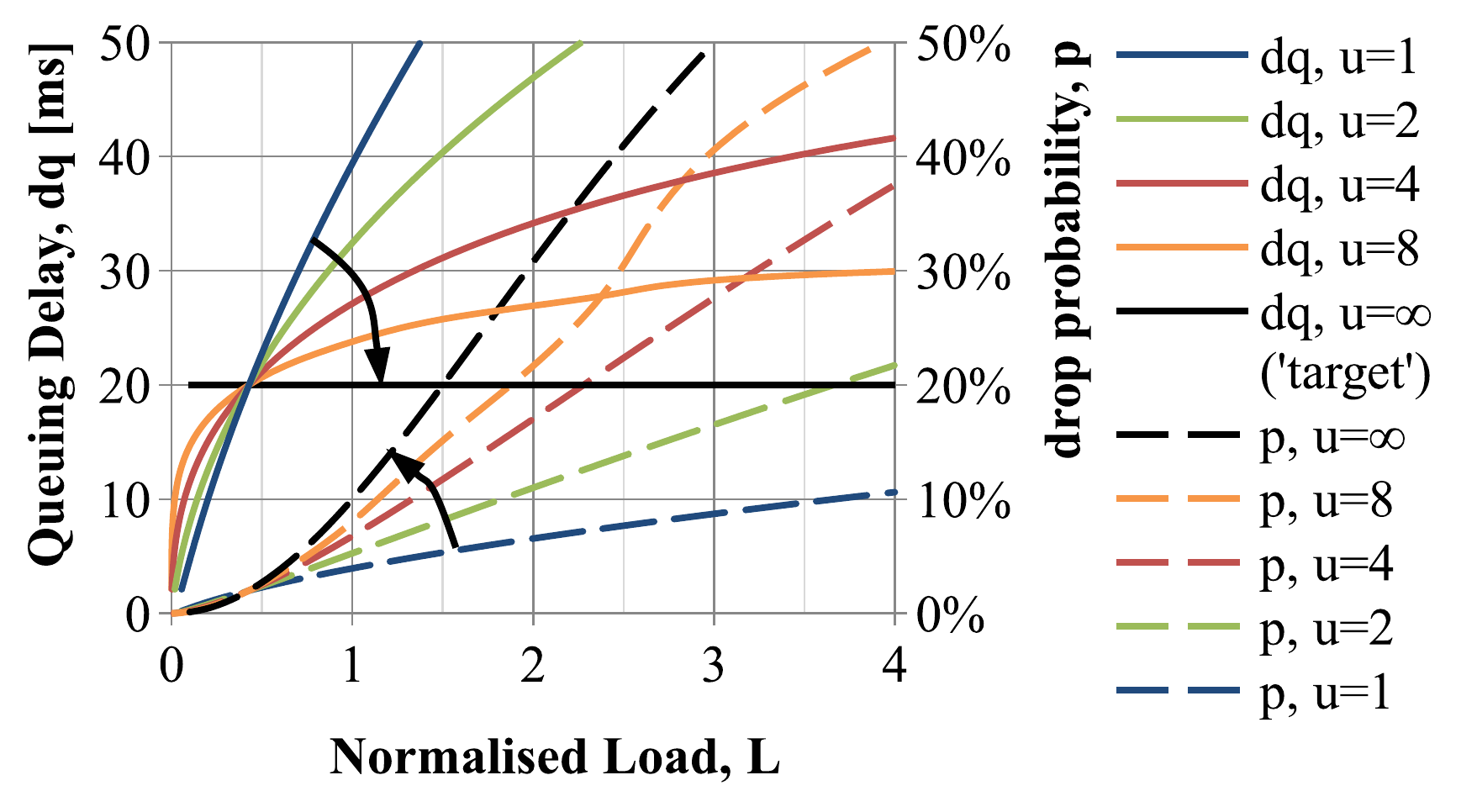}\\
  \caption{Dilemma between Two Impairments: Delay and Loss against 
  Normalised Load; for TCP Reno and the Curvy RED Algorithm with 
  Increasing Curviness, \(u\); a) \(D_R = 20\,\)ms; b) \(D_R = 10\,\)ms}
  \label{fig:credi_dilemma}
\end{figure}

The scaling parameters \(D_q\) of the curves in \autoref{fig:credi_curvy-red-100}
are arranged so that they both pass through (20\,ms, 2\%), which we
call the design point.\footnote{The values are an arbitrary example, not a 
recommendation.} All the curves in
\autoref{fig:credi_curvy-red-5} and \autoref{fig:credi_dilemma} are
arranged to pass through this same design point\footnote{By re-arranging \autoref{eqn:credi_cred} as \(D_q = d_q^* / (p^*)^{1/u}\), where the design point is \((d_q^*, p^*).\)}, which makes them
comparable over the operating region of about 0--40\,ms either side of
this point, shown in \autoref{fig:credi_curvy-red-5}.

In the following, the dependence of both drop and queuing delay on load
will be derived. The approach is relatively simple. The rate of each
flow has to reduce to fit more flows into the same capacity, the number
of flows being a measure of increasing load. The formula for the rate of
a TCP flow is well known; flow rate reduces if either drop or round trip
time (RTT) increases, and RTT increases if queuing delay increases. The
Curvy RED formula constrains the relationship between drop and queuing
delay, so one can be substituted for the other in the TCP formula to
cast it solely in terms of either drop or queuing delay, rather than
both. Therefore either drop or queuing delay can be stated in terms of
the number of flows sharing out the capacity.

\autoref{fig:credi_dilemma} shows the resulting dependence of both
queuing delay (left axis) and drop probability (right axis) on load.
There is a pair of curves for each of a selection of different values of
curviness, \(u\), in the Curvy RED algorithm. The rest of this section
gives the derivation of these curves, and defines the normalised load 
metric used for the horizontal axis.

In practice, traffic is not all TCP and TCP is not all Reno. Nonetheless, 
to gain insight it will be sufficient to assume all flows are TCP Reno. 
Focusing on Reno is justified for bandwidth delay products typical in today's public 
Internet, where TCP Cubic typically operates in a TCP Reno emulation 
mode (e.g.\ with 
20ms round trip time, Cubic remains in Reno mode up to 500Mb/s). The 
following Reno-based analysis would apply for Cubic in Reno mode, except with a 
slightly different constant of proportionality. Even if the dominant congestion 
control were Cubic in its pure Cubic mode, the exponents and constants 
would be different in form, but the structure of the analysis would be the same.

In typical traffic a large proportion of TCP flows are short and
complete while still in slow start. However, again for intuitive
simplicity, only flows in congestion avoidance will be considered. \footnote{If necessary, the traffic can be characterised as an \emph{effective} number, \(n\), of Reno flows in congestion avoidance, for instance by reverse engineering the output of an algorithm designed to measure queue variation \(\nu\) such as ADT~\cite{Stanojevic06:ADT}, i.e.\ \(n = (\mathrm{BDP}/\nu)^2\).}

The load on a link is proportional to the number of simultaneous flows
being transmitted, \(n\). If the capacity of the link (which may vary)
is \(X\) and the mean bit-rate of the flows is \(x\), then:
\begin{equation}
    n = \frac{X}{x}.
    \label{eqn:credi_n1}
\end{equation}

The rate of a TCP flow depends on round trip delay, \(d_R\) and drop
probability, \(p\). A precise but complex formula for this dependence has been derived, but the
simplest model~\cite{Mathis97:TCP_Macro} will suffice for insight
purposes:
\begin{equation}
    x = \frac{Ks}{d_R\sqrt{p}},
    \label{eqn:credi_reno}
\end{equation}
where \(s\) is the maximum segment size (MSS) of TCP and K is \(\sqrt{3/2}\) 
for TCP Reno (or 1.68 for Cubic in Reno mode).

The round trip time, \(d_R\) consists of the base RTT \(D_R\) plus
queuing delay \(d_q\), such that:
\begin{equation}
    d_R = D_R + d_q
    \label{eqn:credi_RTT}
\end{equation}
For a set of flows with different base round trip times, it is sufficient 
to define \(D_R\) as the harmonic mean of the set of base RTTs. \(d_q\) is 
common to all the flows, because they all pass through the same 
queue\footnote{Assuming the common case of a single bottleneck.}, 
so then \(d_R\) is the average RTT including this queuing delay \(d_q\).

Substituting \autoref{eqn:credi_reno} \autoref{eqn:credi_RTT} in
\autoref{eqn:credi_n1}:
\begin{align}
    n &= \frac{X(D_R+d_q)}{Ks}\sqrt{p}.
    \label{eqn:credi_n2}
\end{align}

By substituting from \autoref{eqn:credi_cred} into
\autoref{eqn:credi_n2}, the number of flows can be given as a function
of either queueing delay \(d_q\) or loss probability, \(p\):
\begin{equation}
    n = \frac{X(D_R+d_q)}{Ks}\left(\frac{d_q}{D_q}\right)^{u/2}
    \label{eqn:credi_nq}
\end{equation}
\begin{equation}
    n = \frac{X(D_R + D_q  p^{1/u})p^{1/2}}{Ks}
    \label{eqn:credi_np}
\end{equation}

\autoref{eqn:credi_nq} \& \autoref{eqn:credi_np} are plotted against
normalised load in \autoref{fig:credi_dilemma} for an example set of
Curvy RED algorithms with curviness parameters \(u=1,2,4,8,\infty\),
with base RTT \(D_R=20\,\)ms. \(u=\infty\) represents an algorithm like
PIE or CoDel that attempts to clamp queuing delay to a target value
whatever the load. 

How to read \autoref{fig:credi_dilemma}: For any
one value of \(u\), only two plots apply. Each pair shares a
common colour. The solid line in each pair shows growth in queuing delay (left axis) 
with load and the dashed line shows growth in drop (right axis). For instance,for
\(u=1\), the two blue plots apply, which are uppermost and lowermost. As
the value used for \(u\) increases, the relevant pair of plots to use progress inwards;
down from the top and up from the bottom, like a closing pair of
pincers. 

Normalised load \(L\) is used as the horizontal
axis, because it is proportional to the number of flows \(n\) but scales with
link capacity \(X\), by the following relationship:
\begin{equation}
    L = \frac{K s}{D_R} \frac{n}{X},
    \label{fig:credi_norm_load}
\end{equation}
where \(K s / D_R\)  can be considered constant, at least as long as 
MSS and the typical topology of the Internet are constant. The plot
uses \(s = 1500\)\,B.

It can be seen that, as the curviness parameter is increased, the AQM
pushes harder against growth of queuing delay as load increases.
However, given TCP is utilising the same capacity, it has to cause more
loss (right axis) if it cannot cause more queuing delay (left axis).

In the extreme, infinite curviness represents the intent of AQMs such
as CoDel and PIE that aim to clamp queueing delay to a target value
(the black horizontal straight line). As a consequence, the TCP flows
force loss to rise more quickly with load (the black dashed line). 
The experiments comparing Curvy RED with 
RED, PIE and fq\_CoDel in \cite{DeSchepper15b:DCttH_TR} in a realistic 
broadband testbed verify this analysis, at least for \(u=2\). Indeed the 
high drop levels found with PIE and fq\_CoDel in these experiments 
motivated the analysis in this report.

To reduce losses at high load, the default target delay of fq\_CoDel
probably needs to be increased to a value closer to PIE's target default
of 20\,ms. However then, when load is below the design point, both PIE
and CoDel will tend to allow a small standing queue to build so that the
queue is sufficient to reach the target delay. Whereas Curvy RED should
give lower queuing delay when load is light.

Normalised load is plotted on the horizontal axis to make the plots
independent of capacity as long as individual flows have the same rate.
As an example, we will set the constants to average base RTT, \(D_R =
20\)\,ms and MSS, \(s=1500\)\,B. Then, if the average rate of individual
flows, \(x=1\)\,Mb/s normalised load will be \(L= 1500*8*\sqrt{3/2}
/(0.02*1\mathrm{E}6) = 0.73\), whether there are \(n=4\) flows in
\(X=4\)\,Mb/s of capacity, or \(n=400\) flows in \(X=400\)\,Mb/s.

From \autoref{fig:credi_norm_load} it can be
seen that normalised load depends on the choice of average base delay
used to characterise all the paths, \(D_R\). The horizontal scale of
Figures \ref{fig:credi_dilemma}a) and \ref{fig:credi_dilemma}b) has been
contrived so that a vertical line through them both represents the same
mean rate per flow, \(x\).\footnote{Interestingly, normalised load can
be expresssed as proportionate to the product of number of flows and the ratio of two
delays: \(L = n K D_s/D_R\), where \(D_s\) is the serialisation delay of
a maximum sized segment, because \(D_s = s/X\).} A few selected 
values of rate per flow are shown along the top of  \autoref{fig:credi_dilemma}a).

For instance, if \(D_R=20\)\,ms then \(x\) is 1\,Mb/s when \(L=0.73\),
but if \(D_R=10\)\,ms, \(x\) is 1\,Mb/s when \(L=1.46\). At twice the
normalised load in both plots, \(x\) will halved. As well as this scaling,
there is still residual dependence on \(D_R\) in the shape of the plots
as well, even though normalised load is defined to include average RTT
\(D_R\).

       % Body
% !TeX root = credi_tr.tex
% ================================================================
\section{Invariance with Capacity}\label{sec:credi_capacity}
% ----------------------------------------------------------------
\subsection{Does Flow Aggregation Increase or Decrease the
Queue?}\label{sec:credi_aggr}

% Explain normalisation to design point, and use a more realistic design point in Fig 3.

% Explain why you can only exploit the reduction in queue size due to flow aggregation on very long timescales
%  I.e. the link capacity is designed for a certain no. of flows
%  If there are more than the design intent, queuing delay and drop prob should go up
%  But if the design intent increases (by increasing capacity or reducing load), the design point should reduce
% This effectively means that the queue can self-configure for the 1%ile min flows over hours
%  but if it adapts on the timescale of flow arrivals, it will cause too much drop.

There used to be a rule of thumb that a router buffer should be sized
(in bytes) for the bandwidth delay product. Then, in 2004 Appenzeller et
al.~\cite{Appenzeller04:Sizing_buffers, Ganjali06:Sizing_buffers}
pointed out that for large aggregates \(\mathrm{BDP}/\sqrt{n}\) would be
sufficient. This applies only if
TCP's sawteeth are desynchronized so that the variance of all the
sawteeth shrinks with the square root of the number of flows.\footnote{This square root comes from the 
Central Limit Theorem and has nothing to do with the TCP Reno rate
equation.}

A link for \(100\times\) more flows will typically be sized with
\(100\times\) more capacity, so the queue will drain \(100\times\)
faster. Therefore, if the buffer is sized for
\(\mathrm{BDP}/\sqrt{n}\), queuing delay will be \(\sqrt{n}\mathord\times\) 
\((\mathord=10\times\)) lower.\footnote{Note that this square-root law does not apply indefinitely; only for
medium levels of aggregation. Once the queue size has reduced to about half a dozen
packets, it does not get any smaller with increased
aggregation~\cite{Ganjali06:Sizing_buffers}.}

In the previous section we showed that queuing delay \emph{grows} with
number of flows. How does this reconcile with the idea that buffers 
can \emph{shrink} as more flows are aggregated? 

It is certainly true that, with more flows,
variability of the queue will shrink. So, if curvy RED allows the extra
flows to increase the queue, most of the variation will be at the tail
of the queue in front of a small standing queue at the head. However, without 
more capacity,
the alternative would be for the AQM to introduce more drop to prevent
the additional delay---the dilemma in \autoref{fig:credi_dilemma} cannot
be escaped\footnote{except using ECN ---see \S\,\ref{sec:credi_ECN}}. 

Nonetheless, if the greater number of flows is expected to persist,
capacity can be increased. This creates enough space again for each
flow, so less queuing delay and less loss is needed. For example, if
there are permanently ten times more flows, then provisioning ten times
more capacity will bring \emph{normalised} load back to its original 
value, because normalised load, \(L \propto n/X\).

Note that our definition of normalised load also factors out a change
in MSS. For a given capacity and number of flows,
increasing the MSS increases the normalised load. It may be
counter-intuitive that the same bit-rate in larger packets will
change normalised load. This is an artefact of the 
design of the TCP Reno algorithm (and most other TCP variants), which 
congests the link more (more delay and/or drop) for a larger MSS, 
because it adds one MSS to the load per RTT (the additive increase). 

% ----------------------------------------------------------------
\subsection{AQM Configuration and Scale}\label{sec:credi_aqm_config}

AQM configuration should be invariant with link capacity, but it will
need to change in environments where the expected range of RTTs is
significantly different. Therefore, AQM configurations in a data centre
will be different, not because of high link capacities, but because of
shorter RTTs.

The parameter \(D_q\) represents a delay well outside the normal operating 
region, so it has no intuitive meaning. Therefore it is better to configure the 
AQM against a design point, such as \(d_q =
20\,\mathrm{ms}, p = 2\%\). The operator will need to set this design point
where delay and loss start to become troublesome for the most sensitive
applications (e.g.\ interactive voice). By setting the AQM to pass
through this point, it will ensure a good compromise between delay and
loss. Then the most sensitive application will survive at the highest
possible load, rather than holding one impairment unnecessarily low so
that the other is pushed so high that it breaks the sensitive app.

The ideal curviness of the AQM depends on the dominant congestion
control regime (this memo is written assuming TCP Reno is dominant,
which includes Cubic in Reno mode). For a particular dominant congestion
control, curviness depends only on the best compromise to resolve the
dilemma in \autoref{fig:credi_dilemma}, and can otherwise be assumed
constant.

Having determined an AQM configuration as above, it will be applicable
for all link rates as long as expected round trip times are unchanged.
It would seem that higher flow aggregation would allow queuing delay to
be reduced proportionate to \(1/\sqrt{n}\) without losing utilisation.
However, \emph{the dilemma in \autoref{fig:credi_dilemma} still applies,
so loss probability would increase}. Therefore, once a good balance
between queuing delay and loss has been determined it will be best to
stick with that for any level of aggregation. Then, on a highly
aggregated link, in comparison to an equivalent level of normalised load
on a smaller link with fewer flows, queue variation will be smaller
but it will have to sit behind a larger standing queue. Nonetheless, this is preferable to
changing the AQM scaling parameter to trade more loss for less queuing.

This should help to explain why Vu-Brugier 
\emph{et al}~\cite{Vu-Brugier07:Critique_Buffer_Sizing} found that 
they could not reduce the size of a buffer to \(\mathrm{BDP} / \sqrt{n}\) 
without triggering high volumes of user complaints due to excessive loss levels.

If there are too many flows for the capacity, no amount of AQM
configuration will help. Capacity needs to be appropriately upgraded.
The following formula can be used to determine capacity \(X\) given the
expected number of flows \(n\) and expected average base RTT \(D_R\),
where \(d_q^*\) and \(p^*\) are the values of queuing delay and drop at
the design point, as determined above.
\begin{equation*}
    X = \frac{Kns}{(D_R+d_q^*)\sqrt{p^*}}.
\end{equation*}
Alternatively, the formula can be used to determine the number of flows \(n\)
that a particular capacity \(X\) can reasonably be expected to support.

For a bottleneck link, the only way to take
advantage of the \(1/\sqrt{n}\) reduction in queuing delay due to
aggregation without increasing loss is to over-provide capacity, so
that \(n/X\) reduces as much as \((D_R+d_q^*)\) reduces, thus holding \(p^*\) 
unchanged. 

The necessary degree of overprovisioning can be calculated for the worst
case where the overprovisioned link remains as the bottleneck. In this
case, the design point for queuing delay at low aggregation, \(d_q^*\)
is reduced to \(d_q^*/\sqrt{n}\) at high aggregation, then the
over-provisioning factor should be:
\begin{equation*}
    \frac{X^{\prime}}{X} = \frac{(D_R+d_q^*)}{(D_R+d_q^*/\sqrt{n})}.
\end{equation*}
For example, if average base RTT, \(D_R=20ms\) a link for \(100\times\)
more flows could have the design point for queuing delay reduced from
20\,ms to 2\,ms while keeping the loss design point constant at 2\%.
But only if aggregated capacity is over-provided by \((20+20)/(20+2) =
1.8\), i.e.\ 
\(180\times\) more capacity for \(100\times\) more flows.

If a core link is over-provisioned such that it is rarely the
bottleneck, none of this analaysis applies. Its buffer can be sized at
\(\mathrm{BDP} / \sqrt{n}\) without sacrificing utilisation but accommoding queue
variation due to TCP flows bottlenecked elsewhere, and no regard needs
to be taken of the risk of loss, which will be vanishingly small.
       % Aggregation
% !TeX root = credi_tr.tex
% ================================================================
\section{Escaping the TCP Dilemma with ECN}\label{sec:credi_ECN}

Another insight that can be drawn from this analysis is that the
dilemma in \autoref{fig:credi_dilemma} disappears with ECN. For ECN,
the AQM can mark packets without introducing any impairment. There is
therefore no downside to clamping down queuing delay for ECN-capable 
packets. This proves that it is wrong to treat ECN the same as drop.

When ECN was first standardised~\cite{IETF_RFC3168:ECN_IP_TCP} , it was 
defined as equivalent to drop. Even though earlier proposals had considered 
treating ECN differently, it had to be standardised as equivalent otherwise 
no consensus could be reached on how to define its meaning. 

It is now known that drop probability should be proportional to the square of 
ECN marking, and the constant of proportionality need not be standardised~%
\cite{DeSchepper15b:DCttH_TR}. Therefore it might now be feasible to reach 
consensus on a standard meaning for ECN different from but related to drop.

% ================================================================
\section{Conclusions \& Further Work}\label{sec:credi_Conclusions}

Although Curvy RED seems to be a useful AQM, we are not necessarily
recommending it here. We are merely using the concept of curviness to
draw insights.

The curviness parameter of Curvy RED can be considered to represent the
operator's policy for the tradeoff between delay and loss whenever load
exceeds the intended design point. In contrast, PIE and RED embody a
hard-coded policy, which dictates that holding down delay is paramount,
at the expense of more loss.

Given losses from short interactive flows (e.g.\ Web) cause considerable
delay to session completion, trading less queuing delay for more loss is unlikely to
be the optimal policy to reduce delay. Also, as load increases, it may
lead real-time applications such as conversational video and VoIP to
degrade or fail sooner. Allowing some additional flex in queueing delay
with consequently less increase in loss is likely to give more
favourable performance for a mix of Internet applications.

This dilemma can be escaped by using ECN instead of loss to signal 
congestion. Then queuing delay can be capped and the only consequence 
is higher marking, not higher drop.

The analysis also shows that, once a design point defining an
acceptable queuing delay and loss has been defined, the same
configuration can be used for the AQM at any link rate, but only for a
similar RTT environment. A shallower queuing delay configuration can be
used at high aggregation, but only if the higher loss is acceptable, or
if the capacity is suitably over-provisioned. Formulae for all these
configuration trade-offs have been provided.

We intend to conduct experiments to give advice on a compromise level
of curviness that best protects a range of delay-sensitive and
loss-sensitive applications during high load. It will also be necessary to verify 
the theory for all values of curviness, and for other AQMs without their 
default parameter settings.

Further research is needed to understand how best to average the queue
length, and how best to configure the averaging parameter.

% ================================================================
\section{Acknowledgements}\label{sec:credi_Acks}

\balance%
Thanks to Koen De Schepper of Bell Labs for suggesting to use the
harmonic mean of the base RTTs, rather than assuming equal RTT flows.
Thank you to the following for their review
comments: Wolfram Lautenschl{\"a}ger and the participants in the RITE project, especially, Gorry Fairhurst, Michael Welzl, Anna Brunstr{\"o}m, David Hayes, Nicolas Kuhn and Naeem Khademi.

The author's contribution is part-funded by the European
Community under its Seventh Framework Programme through the Reducing
Internet Transport Latency (RITE) project (ICT-317700). The views
expressed here are solely those of the authors.

       % Aggregation,
% ----------------------------------------------------------------

%\newpage
%\onecolumn%
\addcontentsline{toc}{section}{References}

{\footnotesize%
\bibliography{bob}}

% ----------------------------------------------------------------
%\iftr%
%\clearpage
%\twocolumn%
%\appendix
%\onecolumn%
%\fi%
% ----------------------------------------------------------------
%TODO

\onecolumn%
\addcontentsline{toc}{part}{Document history}
\section*{Document history}

\begin{tabular}{|c|c|c|p{3.5in}|}
 \hline
Version &Date &Author &Details of change \\
 \hline\hline
Draft 00A    &19 May 2015 &Bob Briscoe &First Draft\\\hline%
Draft 00B    &23 May 2015 &Bob Briscoe &Explained normalised load\\\hline%
Draft 00C    &23 May 2015 &Bob Briscoe &Changed notation \& added Aggregation section\\\hline%
Draft 00D   &08 Jun 2015   &Bob Briscoe &Corrected explanation about packet size, plus minor corrections\\\hline%
Draft 00E   &10 Jun 2015   &Bob Briscoe &Cited DCttH.\\\hline%
Issue 01      &25 Jul 2015    &Bob Briscoe &Generalised from equal flows; Included Base RTT in Normalised Load; Changed \(s\) from packet size to MSS. Justified standing queue due to flow aggregation.\\\hline%
 Issue 01A  &26 Jul 2015   &Bob Briscoe &Editorial Corrections.\\\hline%
Issue 01B   &29 Jul 2015   &Bob Briscoe &Clarified applicability and more comprehensive advice on scaling AQM config.\\\hline%
\metaversion &\metadate   &Bob Briscoe &Corrections.\\\hline%
\hline%
\end{tabular}

\end{document}